\documentclass[aps,prb,twocolumn,amsmath,amssymb]{revtex4-1}
\usepackage{subfigure}

\usepackage[normalem]{ulem}
\usepackage{amsmath,amssymb}
\usepackage[bookmarks=true,colorlinks,linkcolor=blue,urlcolor=blue,citecolor=blue]{hyperref}
\usepackage{graphicx,amsmath,amsfonts}
\usepackage[usenames]{color}
\usepackage{epstopdf}
\usepackage{verbatim}
\usepackage{amsthm}
\usepackage{relsize}

\newcommand{\be}{\begin{equation}}
\newcommand{\beq}{\begin{eqnarray}}
\newcommand{\eeq}{\end{eqnarray}}

\def \ua{{\uparrow}}
\def \da{{\downarrow}}
\def \tua{{\tilde\uparrow}}
\def \tda{{\tilde\downarrow}}
\def \be{\begin{equation}}
\def \ee{\end{equation}}
\def \ba{\begin{array}}
\def \ea{\end{array}}
\def \bea{\begin{eqnarray}}
\def \eea{\end{eqnarray}}
\def \nn{\nonumber}

\def \ts{{\tilde{\sigma}}}
\def \ta{{\tilde{\alpha}}}

\def \k{{\kappa}}

\def \a{{\alpha}}
\def \t{{\theta}}

\def \D{{\Delta}}
\def \d{{\delta}}
\def \w{{\omega}}
\def \s{{\sigma}}
\def \S{{\Sigma}}

\def \yd{^\dagger}
\def \av#1{{\langle#1\rangle}}
\def \ket#1{{\,|\,#1\,\rangle\,}}
\def \bra#1{{\,\langle\,#1\,|\,}}

\def \SA {{\tilde{\s},\tilde{\a}}}
\def \UA {{\tilde{\ua},\tilde{\a}}}
\def \DA {{\tilde{\da},\tilde{\a}}}

\begin{document}

\title{Localization and topology protected quantum coherence at the edge of `hot' matter
}
\author{Yasaman Bahri$^1$, Ronen Vosk$^2$, Ehud Altman$^{1,2}$, and Ashvin Vishwanath$^1$\\{\small $^1$\em Department of Physics, University of California, Berkeley, California 94720, USA }\\ {\small $^2$\em Department of Condensed Matter Physics, Weizmann Institute of Science, Rehovot 76100, Israel}}
\begin{abstract}
Topological phases are often characterized by special edge states confined near the boundaries by an energy gap in the bulk. On raising temperature, these edge states are lost in a clean system due to mobile thermal excitations.  Recently, however, it has been established that disorder can localize an isolated many-body system, potentially allowing for a sharply defined topological phase even in a highly excited state. Here we show this to be the case for the topological phase of a  one-dimensional magnet with quenched disorder which features spin one-half excitations at the edges. The time evolution of a simple, highly excited initial state is used to reveal quantum coherent edge spins. In particular, we demonstrate, using theoretical arguments and numerical simulation, the coherent revival of an edge spin over a time scale that grows exponentially larger with system size. This is in sharp contrast to the general expectation that quantum bits strongly coupled to a `hot' many body system will rapidly lose coherence.
 \end{abstract}
\maketitle


\section{Introduction}

Observing coherent quantum phenomena generally requires arranging for some degree of isolation - either physical isolation, to reduce coupling to other degrees of freedom, or by cooling to very low temperatures to freeze out excitations. Therefore, quantum coherence is not expected to survive for long in a general many-body  system placed in an excited state. In particular, quantum bits embedded in such a dynamical system are expected to rapidly decohere. This follows from the assumption of ergodicity -  even a closed many-body system evolving in time under the Schrodinger equation is assumed to approach thermal equilibrium independent of most details of the initial state\cite{ETH1,ETH2,ETH3,ETH4}. In particular, no quantum interference is expected because the outcome would then depend on the initial preparation, in contrast to the expectation for an equilibrium state.

Yet not all many-body systems necessarily thermalize. A well-studied exception to this rule are the so-called
integrable systems. But these are very specific, fine-tuned models whose special properties are lost on adding generic perturbations that are present in a real world realization. A more physically robust mechanism by which quantum coherence can avoid a thermal death is through localization.
Anderson, in his original paper on quantum localization\cite{Anderson1958}, conjectured that closed many-body quantum systems with a sufficiently strong random potential will fail to thermalize. The idea was revived by Basko et. al. \cite{Basko2006}, who gave new theoretical arguments for the stability of the many-body localized (MBL) state in interacting systems decoupled from an external bath. Numerical studies have provided further support for the existence of a non-thermalizing many-body localized state \cite{Oganesyan2007,Pal2010}.

Even if it does not thermalize, a many-body system still has incredibly complex dynamics. With a huge number of correlated degrees of freedom, it is not clear that the system can have any retrievable quantum coherence. Indeed, the time evolution of a many-body system generally involves unbounded growth of the (entanglement) entropy even in the localized state\cite{Znidaric2008,Bardarson2012,Vosk2013,Serbyn2013}, which may suggest irreversible loss of information. To measure and utilize quantum coherence will require at least one addressable degree of freedom that is effectively decoupled from all the rest. Again, this appears to require immense fine tuning.
\begin{figure}
\includegraphics[width=\linewidth]{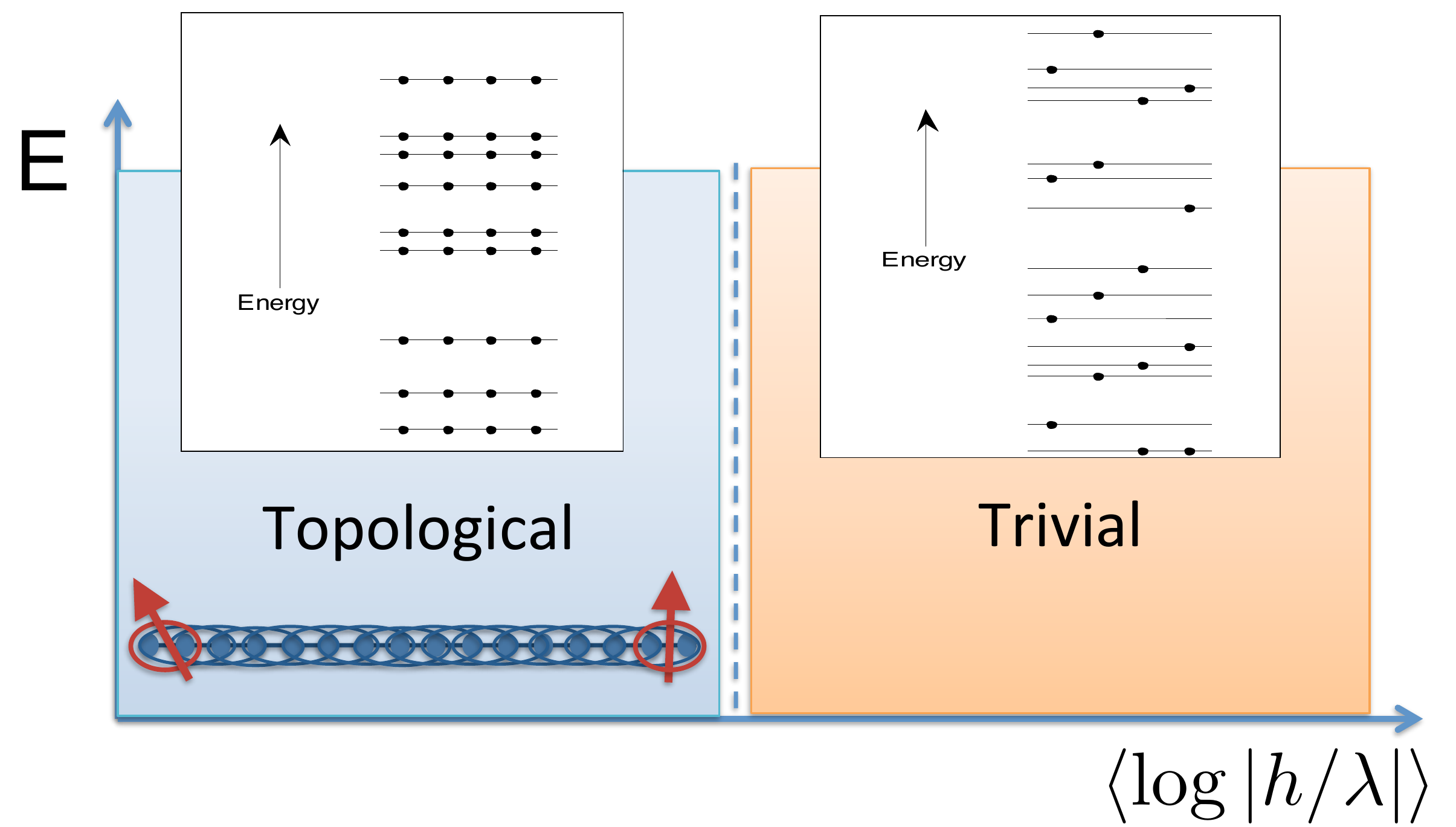}
\caption{Schematic phase diagram of the model in Eqn. \ref{model}. Vertical axis is energy density such that the maximum corresponds to infinite temperature in an equilibrium system. Horizontal axis is a measure of the typical strength of the transverse field to the three-spin term. With strong disorder, all eigenstates are localized but change character from the left to right end. In contrast to the trivial phase (right), the topological phase (left) is characterized by quadrupling of energy levels (horizontal lines) sufficiently deep in this phase. Solid dots are used to highlight the levels and the degeneracy; they are displaced along the  horizontal direction for visibility. Spectrum is shown close to $E=0$ in model Eqn. \ref{model}, with parameter standard deviations $\sigma_V=0.1$ and  $(\sigma_\lambda,\sigma_h)=(1.0,\,0.05)$ on the left and $(0.1,\,1.0)$ on the right. }\label{fig:pd}
\end{figure}

Nonetheless, here we show that a MBL state with a certain topological character can guarantee a dynamically decoupled degree of freedom at the edge, which therefore retains its quantum coherence. This is a result of having topological properties that persist to arbitrarily high-energy eigenstates.

In the more familiar context of ground state topological phases, edge modes are protected by a bulk gap and the symmetries of the system\cite{Gu2009,Chen2011a,Pollmann2012,Turner2011,Fidkowski2011,Schuch2011}.
An example relevant to our discussion is the antiferromagnetic spin-1 chain, which has an energy gap to bulk excitations\cite{Haldane1983} but which, remarkably, hosts a free spin-1/2 state at the edge \cite{Kennedy1987}. The fractional edge spin is a manifestation of a topological property of the ground state, which survives as long as certain spin rotation symmetries are present. Hence this defines a Symmetry Protected Topological (SPT) phase. However, in a clean system at any finite temperature the edge spins mix with delocalized bulk states, and no sharp definition of this phase remains. In the model system we present below, a similar edge spin-1/2 quantum bit appears, but it is preserved even on `heating' the bulk due to localization of the bulk modes.  Here, in the absence of thermal equilibrium, the analog of a high-temperature state is one with an energy that significantly exceeds the ground state, so that even in the thermodynamic limit a finite excitation energy density (energy per spin) remains. The coherence of the edge spin and its robustness to generic interactions in the system is clearly demonstrated below using a combination of theoretical arguments and numerical simulation of the quantum dynamics using exact diagonalization of a spin chain with quenched disorder. Irreversible decoherence occurs only after a time exponentially long in the system size, indicating the q-bit can maintain perfect coherence in the infinite system limit.

\section{Model}
To demonstrate the existence of edge states at high energy densities and to show how they can be manipulated coherently, we shall work with the following spin chain Hamiltonian
\be
H=\sum_i\left [ \lambda_i\sigma_{i-1}^z\s_i^x\s_{i+1}^z + h_i\s^x_i+V_i\s^x_i\s^x_{i+1} \right ]
\label{model}
\ee
Here $\s^\a_i$ represent Pauli matrices at site $i$ of a one dimensional chain and the parameters $\lambda_i,h_i$ and $V_i$ are independent random variables. Note, there are two $Z_2$ symmetries corresponding to $\sigma^{z,y}_a \rightarrow -\sigma^{z,y}_a$ independently on the even and odd numbered sites. We note that the first two terms of (\ref{model}) can be mapped to a non-interacting (Majorana) fermion model using a Jordan-Wigner transformation \cite{Kopp2005}. The last term $V_i$ adds interactions between the fermions, making the model generic.

There are two simple limits in which the Hamiltonian (\ref{model}) can be trivially diagonalized, which represent two distinct dynamical phases. First, if the only terms in (\ref{model}) are the local fields $h_i$, then the system is in a trivial localized state. All eigenstates are then simple product states of the spins in the $\s^x$ basis. The other limit, in which only the three-spin terms $\lambda_i$ are non-zero, represents a non-trivial localized phase, analogous to the Haldane phase of the spin-1 chain, with the $Z_2\times Z_2$ symmetry mentioned above playing the role of spin rotation symmetry. In this phase, all the eigenstates are cluster states with distinct topological properties that are manifested by the presence of edge states. Crucially, the distinction between the two types of spectra can persist away from the simple limit and in the presence of interactions $V_i$ due to the localized nature of the state. Such protection of quantum order by many-body localization was discussed in Ref. [\onlinecite{Huse2013, Bauer2013}].

Fig. \ref{fig:pd} gives a schematic view of the phase diagram showing the transition from the trivial to the topological eigenstates of the model (\ref{model}) with decreasing transverse field $h_i$. The transition between the two localized phases can occur through a direct transition described by an infinite randomness dynamical critical point\cite{Vosk2013a,Pekker2013}, while an intervening delocalized phase appears unlikely\cite{Huse2013} . In the sections below we describe the non-trivial dynamics, which ensues when the system is in the topological phase, stressing, in particular, the behavior of the protected edge modes.

\section{Results}
\begin{figure*}[t]

    \subfigure{(a)} {\includegraphics[width=0.7\linewidth]{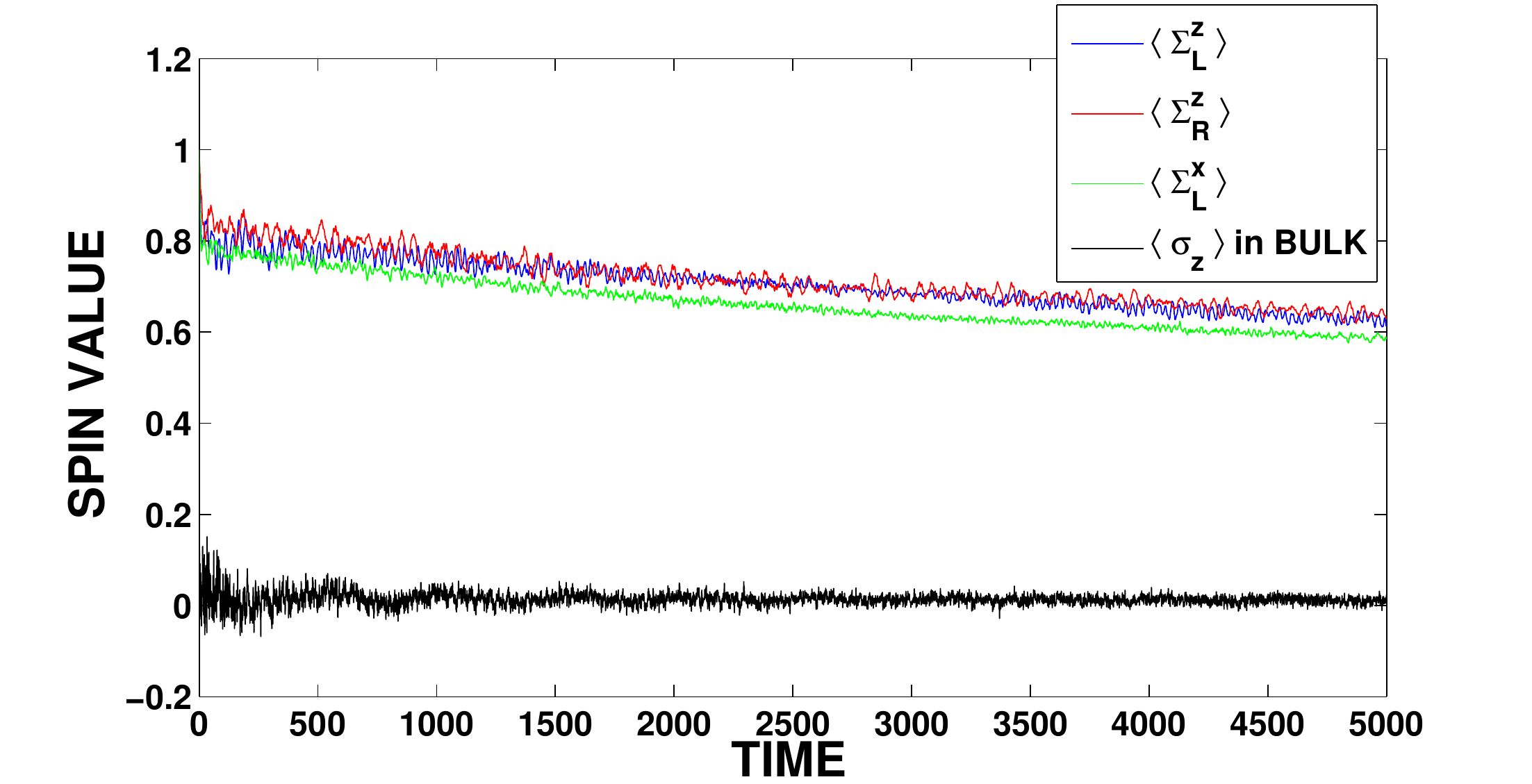} }\\
   \subfigure{(b)} {\includegraphics[width=0.5\linewidth]{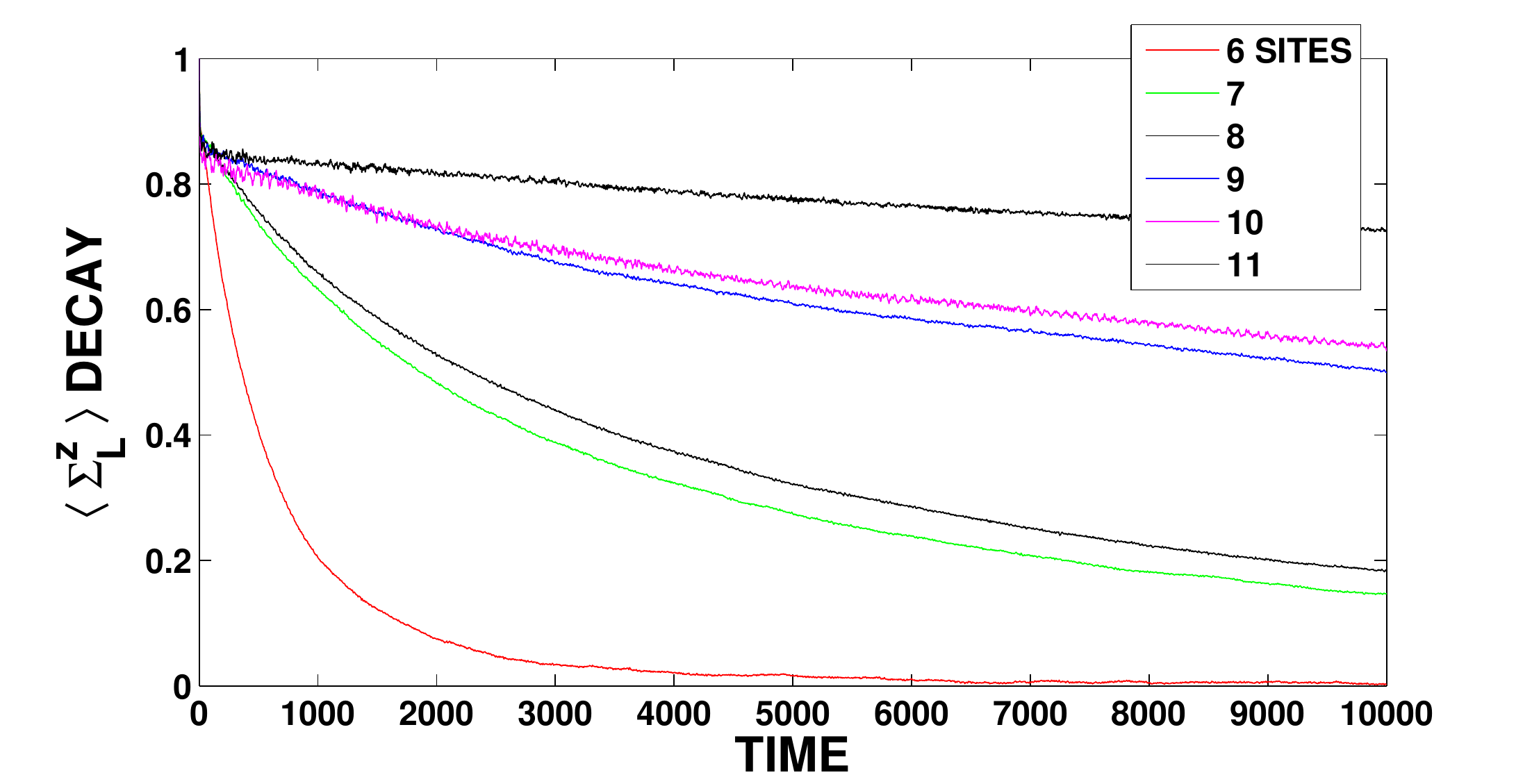} }
    \subfigure{(c)} { \includegraphics[width=0.4\linewidth]{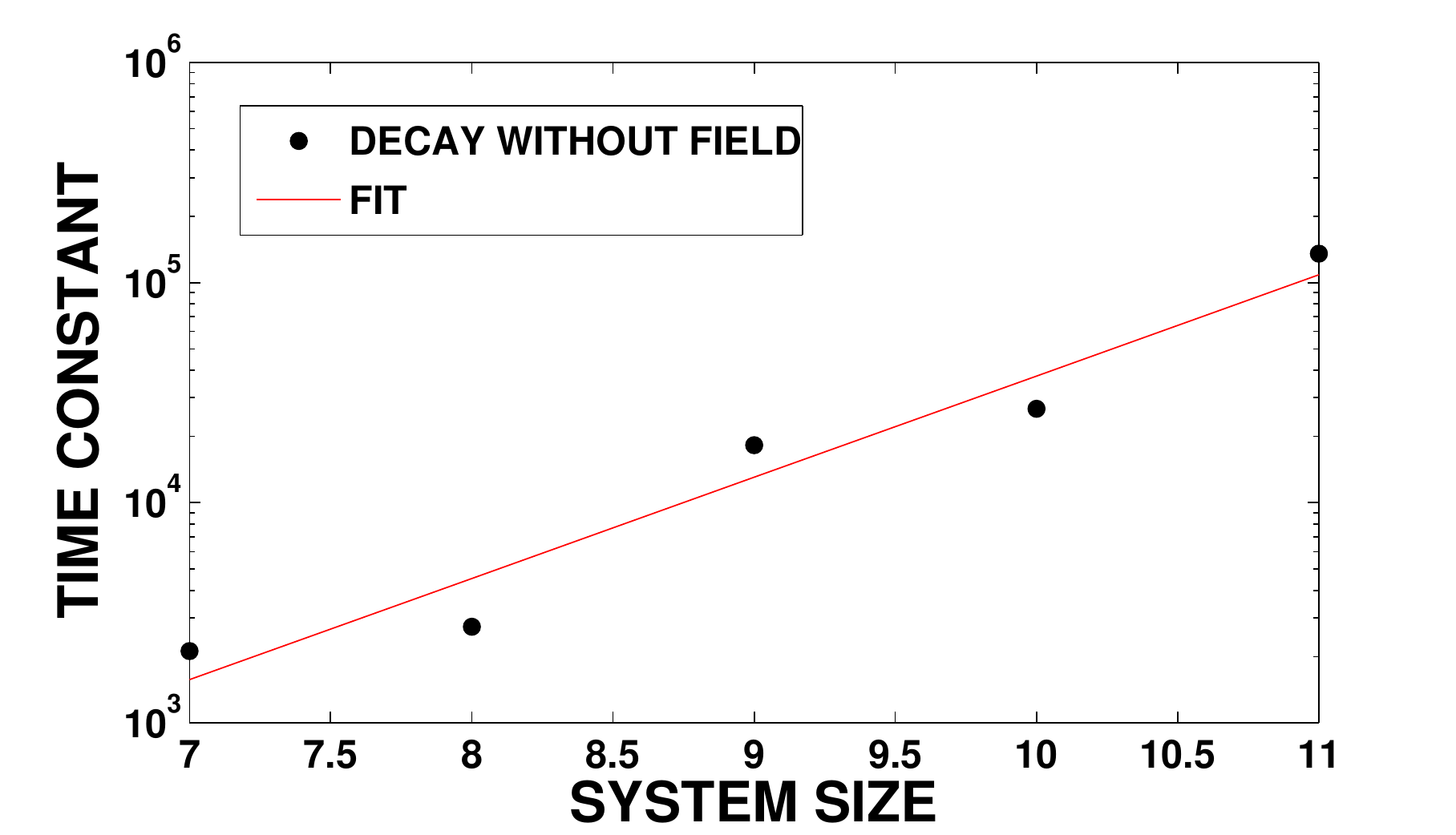}}
\caption{(a)
Decay of an initial state with all spins polarized along $z$ in the topological phase. Disorder averaged $\sigma_z$ at the ends (labeled $\Sigma^z_{L,R}$) and near the center of the ten-site chain show the rapid decay of spin in the bulk but long-lived edge spins. For a different initial state,  $\Sigma^{x}_{L}$ will persist instead, as shown, demonstrating the quantum nature of the edge spin. (Drop from initial +1 value not visible on this time axis scale). (b) The disorder averaged edge spin decay for different system sizes $L$ (c) shows an exponential dependence of the time constant $T_0(L)$ on system size, indicating that the long-time decay of the edge spin originates from coupling to the opposite edge spin. Parameters used had zero mean and standard deviations were $(\sigma_\lambda,\sigma_V,\,\sigma_h)=(1.0,\,0.1,\, 0.05)$. }\label{fig:result1}
\end{figure*}
\subsection{Idealized cluster model}
To understand the dynamics in the Hamiltonian (\ref{model}) and the nature of the eigenstates, consider first the simplified version  having only the first term of (\ref{model}), that is, $H_0=\sum_i\lambda_iK_i$, where $K_i= \sigma_{i-1}^z\s_i^x\s_{i+1}^z$ are mutually commuting operators called stabilizers. Since this model is a sum of stabilizers, the eigenstates are mutual eigenstates of all the $K_i$'s labeled by the respective eigenvalues $\pm 1$.
These states are called cluster states in the quantum information literature. Cluster states have non-trivial entanglement properties since the $K_i$ are spatially overlapping. In fact, they are an example of symmetry protected topological states\cite{Gu2009,Chen2011a,Pollmann2012} with $Z_2\times Z_2$ symmetry\cite{Smacchia2011}. The topological nature of the cluster states is encapsulated in a string order parameter~\cite{Bahri2013} given by $O_{\text{st}}(i,j)=\av{\sigma^z_i\sigma^y_{i+1}\left(\prod_{k=i+2}^{j-2} \sigma^x_{k}\right)\s^y_{j-1}\s^z_j}$, assuming the symmetry remains unbroken. In a specific eigenstate and disorder realization, the string order parameter takes random values $O_{\text{st}}(i,j)=\pm 1$.  Hence we can form the non-local analogue of the Edwards-Anderson glass order parameter $\Psi_{sg}=\overline{O_{\text{st}}^2}$, which gains a non-zero expectation value in the topological glass. This order is found in eigenstates of $H_0$ at all energies.

Another manifestation of topological order that can be seen is in the entanglement spectrum \cite{Pollmann2010}. If we pick an eigenstate of the system with periodic boundary conditions, for instance, and consider the reduced density matrix obtained by tracing over half of the system, the resulting entanglement spectrum (eigenvalues of the density matrix) will be four-fold degenerate in the topological phase due to the spin one-half at each end of the cut\cite{Supplementary}.

The non-trivial structure of the cluster states gives rise to edge modes that behave as free spin-1/2 particles. This is easily seen by constructing a spin-1/2 algebra of edge operators: $\S^x_L=\s^x_1\s^z_2\,,\S^y_L=\s^y_1\s^z_2,\,\S^z_L=\s^z_1$ on the left edge and similarly on the right edge. We say that these edge degrees of freedom behave as free spins because $H_0$ commutes with them. Note that the $\Sigma_L$ operators cannot individually appear in the Hamiltonian since they break the underlying $Z_2\times Z_2$ symmetry.
As an example of the dynamics in this model, we can consider the time evolution starting from an initial product state with the first spin polarized along $x$ and all the rest polarized along $z$. In this state, the left edge q-bit is oriented along $\S^x_L$ and will remain in this state independent of the complicated dynamics of the bulk spins.


\begin{figure*}[t]
\includegraphics[width=0.9\linewidth]{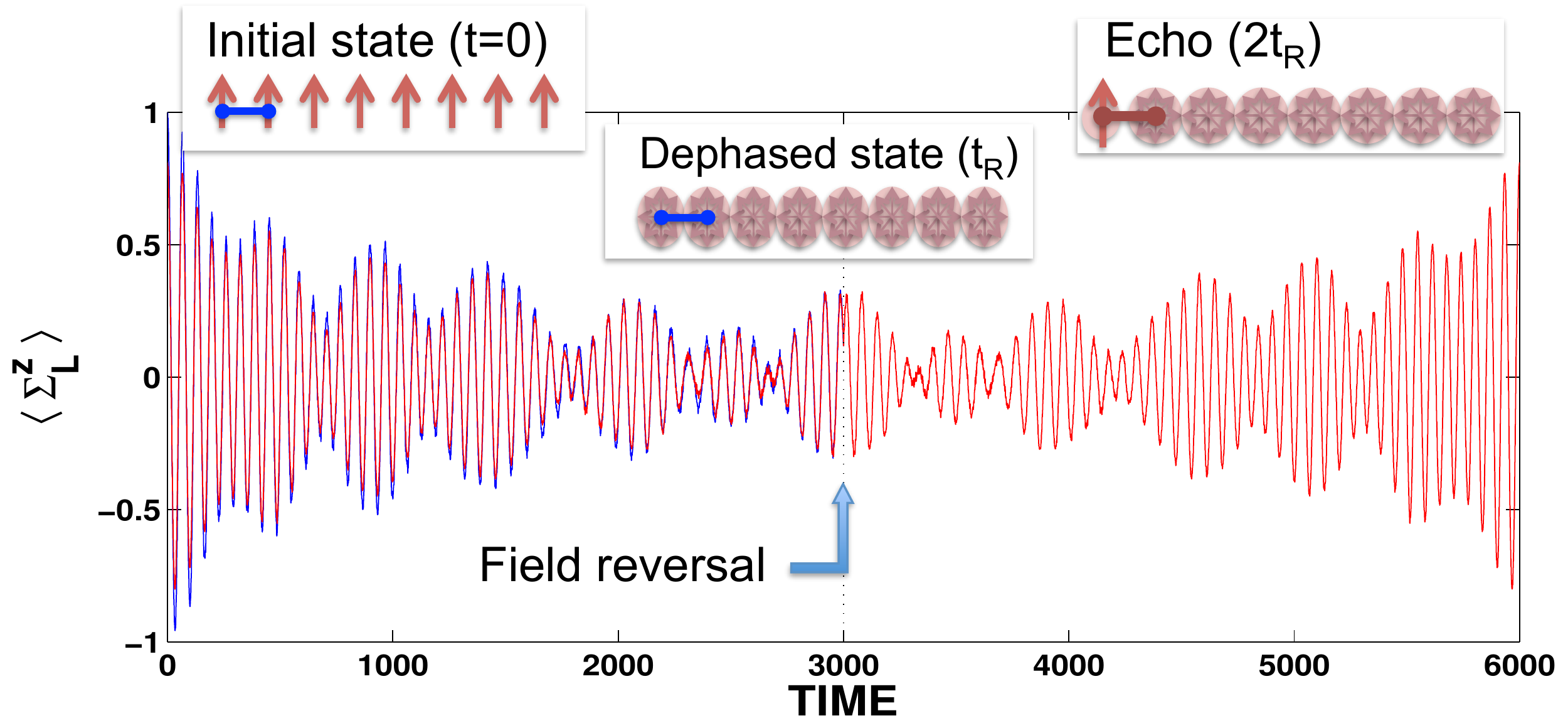}
\caption{The system is prepared with all spins polarized along $z$. Hence, the initial expectation value of the edge spin $\av{\S^z_L}=1$. The state is evolved with the Hamiltonian in the topological phase and an edge field along $\S^x_L$: ~$g \s^x_1 \s^z_2=g\S^x_L$. The edge spin precesses and decays with a length independent time constant ($T^*_2$). However, reversing the edge field $g\to-g$ at time $t=t_R$ leads to a spin echo at time $t=2t_R$, which recovers much of the lost amplitude. The oscillations of the edge spin $\av{\S^z_L}$ are shown here for a particular disorder realization before (blue) and after (red) the field reversal. The red trace is reflected about the time axis to show retracing of the edge spin in the dynamics. }
\label{fig:echo-scheme}
\end{figure*}

\subsection{Effect of interactions}
We now consider adding to the Hamiltonian $H_0$ generic perturbations such as $h_i$ and $V_i$ in (\ref{model}) which destroy its integrability but preserve the $Z_2\times Z_2$ symmetry of the problem. An important simplification occurs in the limit of strong disorder (relevant to the models discussed here), when eigenstates at all energies are localized. Then, one can identify integrals of motion in a many-body localized state\cite{Vosk2013,Serbyn2013a,Huse2013a} that take on a quasi-local character. This will provide a powerful conceptual tool to also describe the localized topological phase and interpret numerical results for the dynamics.


The eigenstates of the toy model discussed above were the strictly localized cluster states. Assuming localization persists on adding the interaction terms, the local stabilizers $K_i$ will map to a set of commuting integrals of motion
 $\tilde{K}_i$ that are still local up to an exponentially decaying tail. Like the original stabilizers $K_i$, each $\tilde{K}_i$ is one projection of a spin-1/2 variable. We term such operators quasi-local. Similarly we will have modified quasi-local left edge operators $\tilde{\S}^\a_L$ that remain decoupled from the dynamical (bulk) Hilbert space in a semi-infinite system.

Experiments can have direct access only to simple, strictly local operators such as the original edge operators $\Sigma^\a_L$ and not to the exact decoupled operators $\tilde{\S}^\a_L$. The crucial point is that if the system is in the localized phase, the measurable quantity $\S^\a_L$ has an overlap with the true edge operator. That is, we can write the original edge operator as a sum of a left edge contribution and a bulk contribution
$
\S^\a_L=Z_\a\,\tilde{\S}^\a_L+ c_\a\,  \tilde{O}^\a_B.
$
If we start the dynamics with $\S^\a_L$ fully polarized in the $\a$ direction, then a polarization of $Z_\a$ will remain after arbitrarily long time evolution.
If, on the other hand, the system is finite, the non-local tails of the new edge operators can lead to interactions between the two edges and the bulk of the system through terms of the form
\be
H_{\text{edge}}= J^{\a\beta}_m\tilde{\S}^\a_L \tilde B_m\tilde{\S}^\beta_R.
\ee
A large number of non-local bulk operators $B_m$ are possible and the respective coefficients are all exponentially small in the system size. The different terms give rise to oscillations at a large number of frequencies, which will in turn lead to decay of the edge spin over a time $\tau_N$ that scales exponentially in the system size N.

Results from numerical diagonalization of the model (\ref{model}) with up to 11 spins are shown in Fig. \ref{fig:result1}.
In one set of calculations, done in a parameter regime deep in the topological state, the system is initialized in a product state with all spins polarized along $\sigma^z$, so that the bare edge operator $\S^z_L$ has a definite value $+1$.
In another set of calculations, we retain the initial conditions above except  for the first spin which is polarized along $\s^x$ so that $\S^x_L$ has a definite value. In both cases the spin expectation value first drops to a smaller nonvanishing value, which later decays to zero on a much longer time scale which we label $T_0$ (Fig. \ref{fig:result1}a). Analysis of $T_0$ as a function of system size confirms the exponential size dependence discussed above. In contrast, the spin expectation value in the bulk of the system always decays rapidly to zero. In the trivial phase ($\av{\log|h/\lambda|}$ larger than a critical value), a uniform initial polarization of $\sigma^z$ decays both in the bulk and at the edge\cite{Supplementary}. Note that while $\langle \sigma^{x}_i \rangle$ may persist in the trivial phase both in the bulk and on the edge, other spin components are not conserved and hence there is no preserved \emph{quantum} spin.

\subsection{Quantum Coherent Dynamics and Spin Echo}

The fact that all components of the topological edge states are conserved in the dynamics suggests that this degree of freedom can indeed store a quantum bit of information. Can it also be manipulated coherently as a q-bit?
To answer this question consider coupling the addressable edge spin to a local field $H_{\text{ext}}=g\S^x_L$.  Since we are in the MBL phase, there is still a constant of motion $\tilde{\S}^x_L$ directly related to $\S^x_L$. But because the field breaks the protecting $Z_2\times Z_2$ symmetry, it induces coupling between the edge operator and the bulk
\be
g\Sigma_L^x=g Z \tilde{\Sigma}_L^x+ g \sum_n c_\a \tilde{\Sigma}^{\a}_L \tilde{B}_\a
\label{edge-int}
\ee
Now instead of observing oscillations at a single frequency $\w=g$ as in the idealized cluster model, the edge spin should precess with many frequencies, leading to decoherence. The decoherence time is independent of system size, set only by the magnitude of the field $g$ and the intrinsic interactions that determine the coefficients $c_n$. At a glance, such dephasing with an infinite number of intrinsic modes appears deadly to the quantum coherence of the q-bit. However, note that all the terms causing the the dephasing of the edge modes are proportional to the external field $g$, or more generally to an odd power of it\cite{Supplementary}. Hence by reversing that field, all oscillations terms are reversed, leading to a spin echo. The overlap between initial and final spins is less than unity but finite and expected to persist to infinite time in the thermodynamic limit. In a finite-size system, there is an additional imperfection due to the field-independent coupling between the two edges which is exponentially small in the system size.

\begin{figure}[b]
 \includegraphics[width=1.0\linewidth]{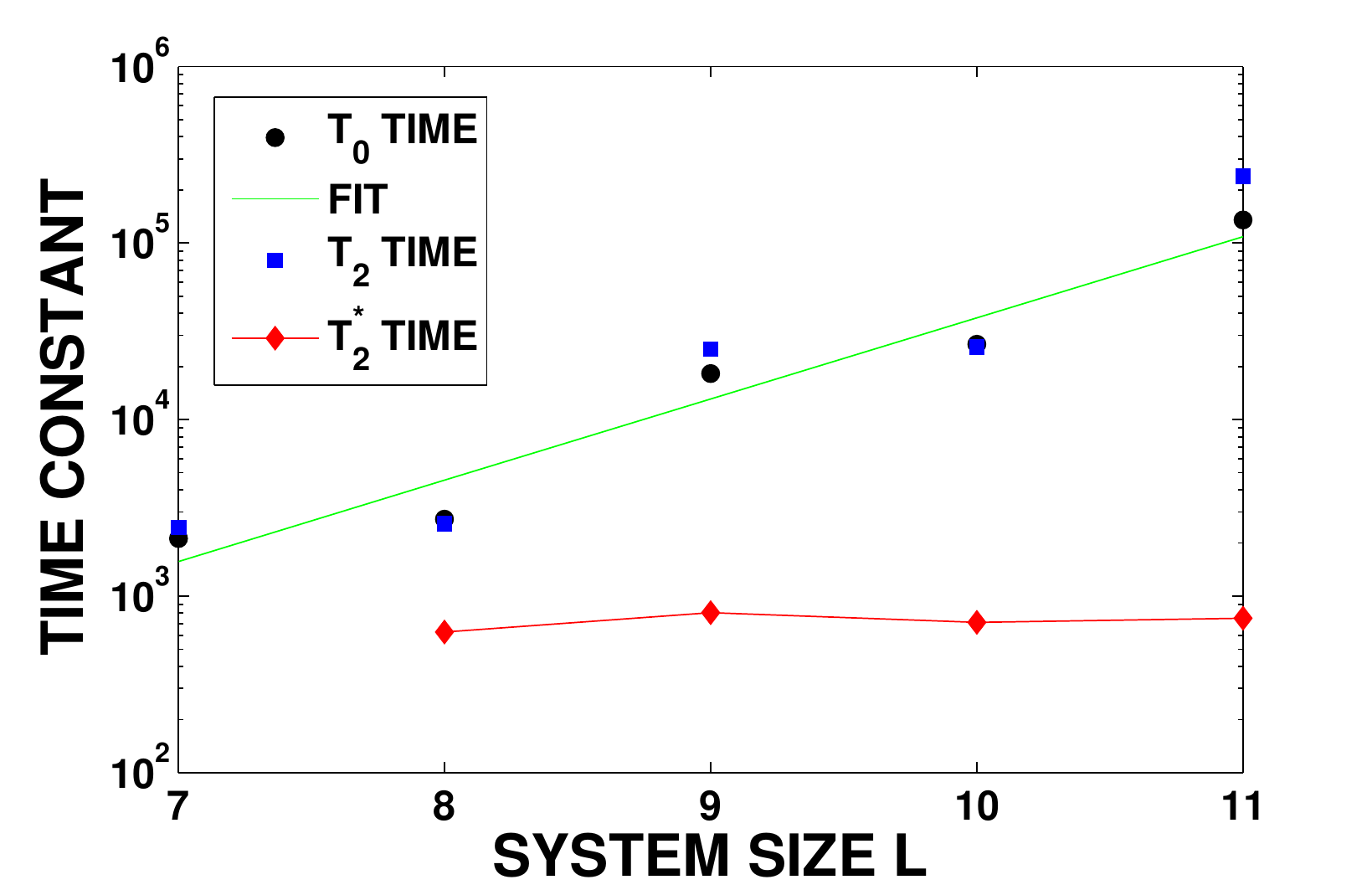}
\caption{Scaling of all the time constants with system size. The zero field decay time $T_0$  and spin echo decay time $T_2$ are nearly identical and scale exponentially with system size, whereas the oscillations decay time $T_2^*$ is size independent.
Parameters used had zero mean and standard deviations were $(\sigma_\lambda,\sigma_V,\,\sigma_h)=(1.0,\,0.1,\, 0.05)$. }
\label{fig:result2}
\end{figure}

To numerically test the degree of edge coherence we computed the time evolution of the edge spin in the model (\ref{model}) supplemented by a local field $g\Sigma_L^x$ applied at one end of a chain.
Fig. \ref{fig:echo-scheme} shows an example of the dynamics in a specific disorder realization for an eight-site chain, starting from a definite value of the left edge $\S^z_L$. As expected we see oscillations of $\av{\S^z_L}$ decaying on a time scale that is independent of the system size (see [\onlinecite{Supplementary}]). This is a dephasing time $T^*_2$ in nuclear magnetic resonance (NMR) terminology. At time $t_R$ we reverse the field $g$ and observe a near perfect retracing of the oscillations. The edge spin echo at $2 t_R$ degrades slowly with increasing the reversal time $t_R$, from which we can extract a relaxation time, called  $T_2$ by analogy with the NMR terminology. It is remarkable that here $T_2$ increases exponentially with growing system size, as shown in Fig. \ref{fig:result2}. This implies that in the large size limit, the edge q-bit can maintain truly perfect coherence in spite of it being embedded in an interacting many-body system with high energy density.

It is interesting to note that in spite of the numerous integrals of motion the bulk of the system reaches, for all practical purposes, an infinite temperature. This is because the initial product states we prescribed do not constrain the value of those integrals of motion. The edge q-bit on the other hand remains `cold' due to the dynamical decoupling from the bulk.

\section{Conclusions}

We have shown that a topological edge state can survive as a coherent degree of freedom at arbitrarily high energies due to localization of the bulk modes. This phenomena opens the way to investigations of topological effects in dynamics. Systems of ultra-cold atoms and ions, which quite naturally fulfill the requirement of being decoupled from an external bath, are a platform for potential realization. Furthermore, the dynamical signatures described here begin with simple initial states. A literal realization of model (\ref{model}) with trapped ions can utilize techniques demonstrated in Ref. [\onlinecite{Lanyon2011}] to generate the three-spin interaction. We stress, however, that three-spin interactions are by no means a fundamental requirement for establishing the strongly disordered topological phase. While the simplest realization of a 1D topological phase - the S=1 antiferromagnet - transitions into a random singlet phase with strong disorder\cite{Hyman1997,Monthus1997}, we believe that other models with two-body interactions that realize this phase can be found. Finally, the topological states of the spin chain (\ref{model}) are related to a much wider classification of symmetry protected topological ground states\cite{Gu2009, Chen2011a, Pollmann2012,Turner2011,Fidkowski2011, Schuch2011}. An interesting question for future investigation is whether the interplay of localization and topology in the higher dimensional members of this classification \cite{Chen2011, Chen2012,Levin2012,Lu2012a,Vishwanath2013,Metlitski2013} can help protect the edge states from mixing with the bulk and thereby lead to novel dynamical phenomena at high energy densities.

\section{Acknowledgements}

We acknowledge illuminating discussions with D. Huse and D. Budker. This work was supported by the ISF (EA), Minerva foundation (EA), the ERC under the UQUAM project (EA),  NSF GRFP under Grant No. DGE 1106400 (YB),  NSF DMR 0645691 (AV) and a Simons Fellowship (AV).  E. A. acknowledges the hospitality of the Miller
institute of Basic research in Science and the Aspen Center for Physics under NSF Grant \# 1066293 for hospitality during the writing of this paper.

\bibliography{mblspt}

\newpage

\appendix

\newpage
{\center \Large{\bf Supplementary Material}}
\vspace{6mm}

In this supplement we give more details of the calculations and additional numerical results. We also present typical entanglement spectra corresponding to high energy states and compare to those in the trivial state.

\section{Length dependent decay of the edge spin at zero field -- $T_0$ time}
\begin{figure}[tbh]
\includegraphics[width=1.0\linewidth]{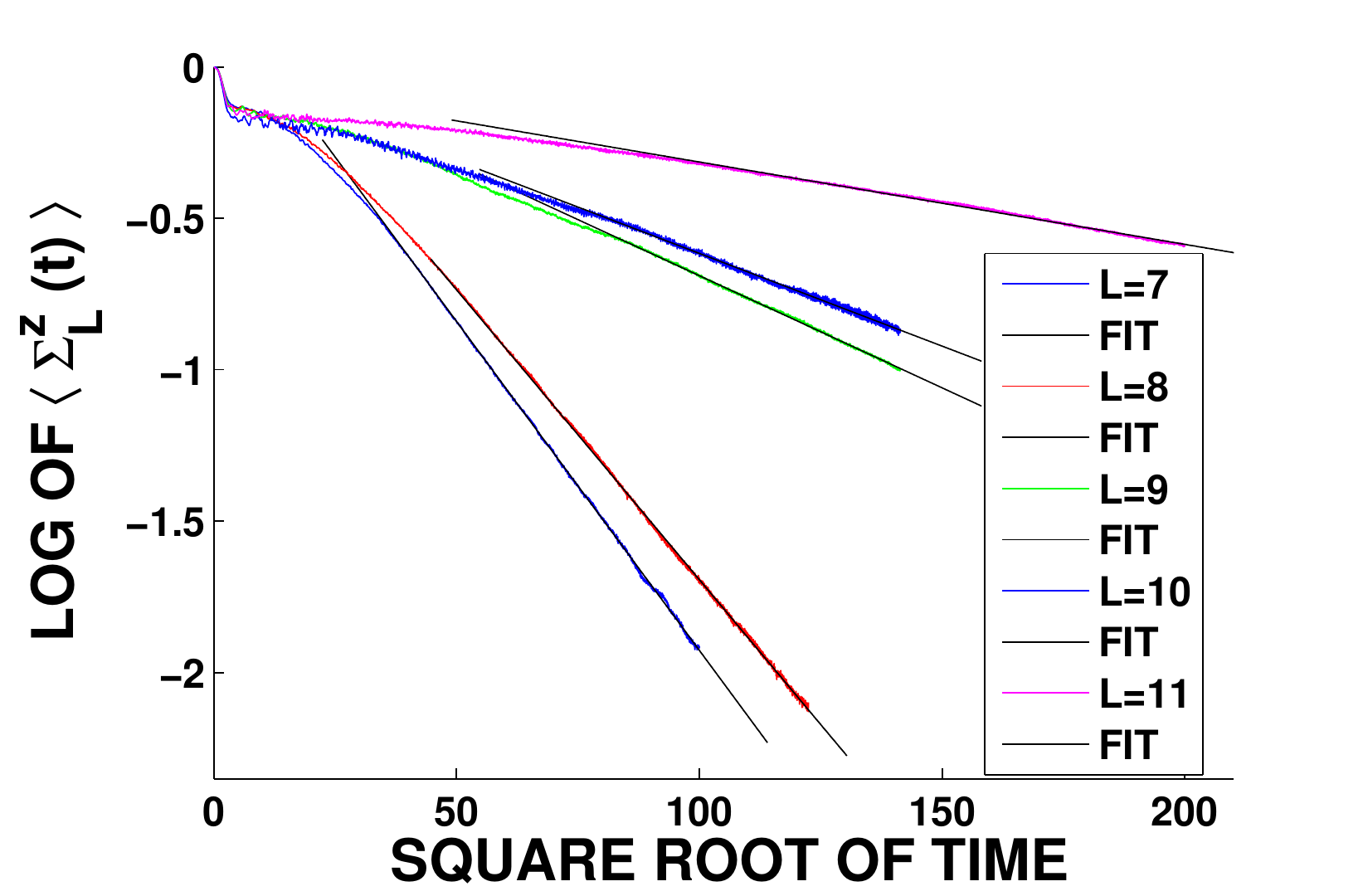}
\caption{Fits of the decay of the edge spin $\av{\S^z_L}$ in zero field.}\label{fig:T0}
\end{figure}

For the numerical calculations we use the hamiltonian
\be
H=\sum_{i} \lambda_{i} \sigma^{z}_{i} \sigma^{x}_{i+1} \sigma^{z}_{i+2} + V_{i} \sigma^{x}_{i} \sigma^{x}_{i+1} + h_{i} \sigma^{x}_{i}
\ee
with open boundary conditions and parameters following normal distributions with zero mean and standard deviations $(1,0.1,0.05)$ in $\lambda, V, h$ respectively. Disorder realizations were selected to have characteristics closely matching the mentioned parameters per sample (within $0.01$ tolerance). Exact diagonalization was used on systems with 6-11 sites. We computed $\left[ \langle \S^z_L (t) \rangle \right]_{dis}$, averaging over disorder and the two samples edges to get smooth curves, and found it fit well to a stretched exponential $C_{0} \exp\left({-\sqrt{\frac{t}{T_{0}}}}\right)$ at longer times (see fits in Fig. \ref{fig:T0}). We used, for L=6-11, $\{ 8000,2000,6400,1900,235,290 \}$ disorder realizations, respectively.

\begin{figure}[tbh]
\includegraphics[width=1.0\linewidth]{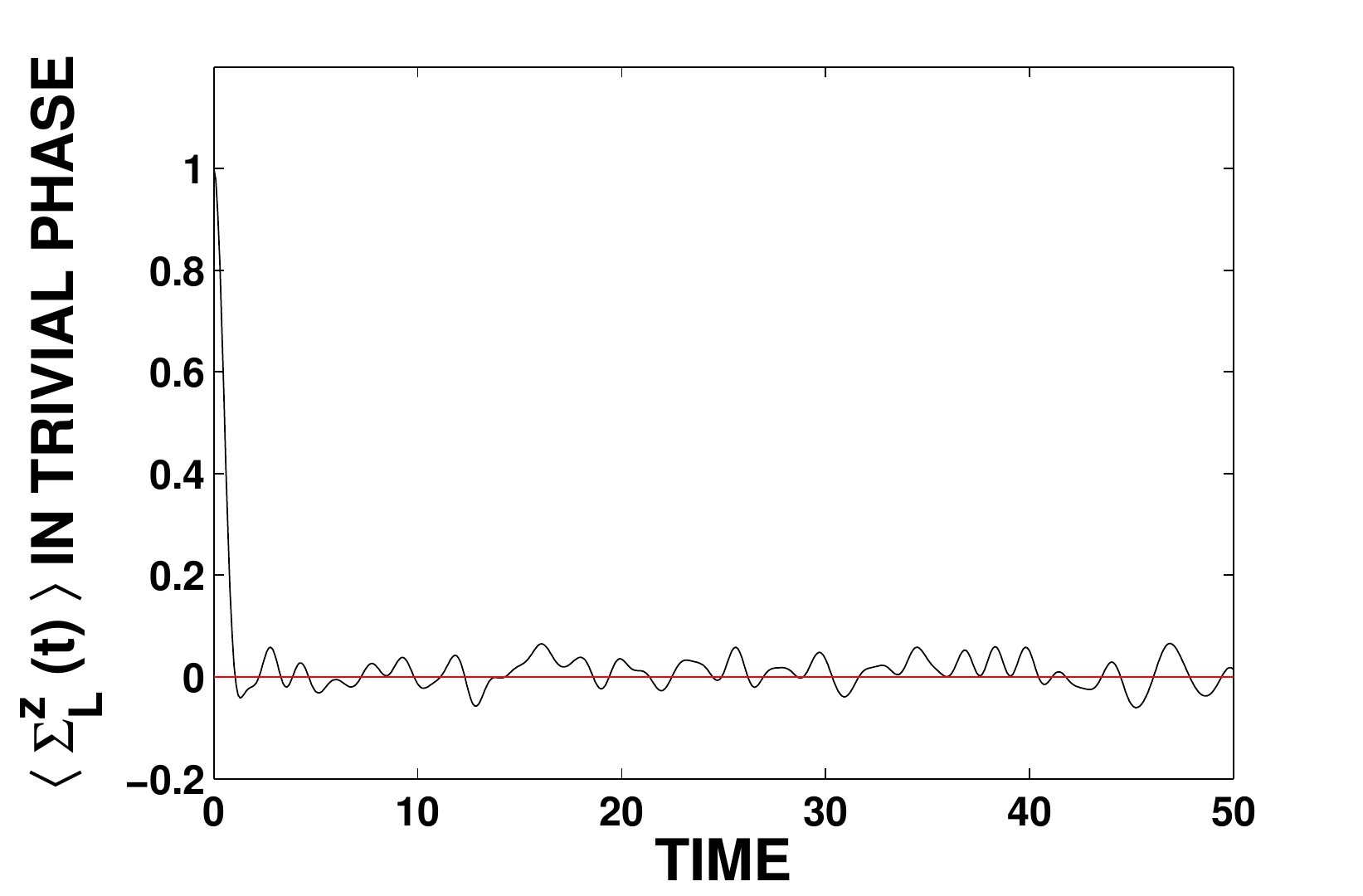}
\caption{Rapid decay of the edge spin in the trivial phase (L=8).}\label{fig:T0trivial}
\end{figure}

In the main text we demonstrated the saturation of the edge-state spin expectation values $\av{\S^\a_L}$ in the thermodynamic limit. We contrast this here to the rapid decay of the edge degrees of freedom in the trivial phase. Fig. \ref{fig:T0trivial} shows the expectation value of the left most spin $\s^z_1$, which in the topological state has a finite overlap with the actual edge operator.

\section{Decay of oscillations in an edge field -- $T_2^*$ time }

We considered the Hamiltonian $H_{0} + B \S^x_L$ with B=0.05 for system sizes 8-11. We chose the value $B=0.05$ because: (i) for much smaller fields it is more difficult to separate the field-dependent time scale $T_2^*$ from the size dependent $T_2$ for the system sizes we can access; (ii) much larger fields move the edge inwards and thus reduce the overlap between the local operators $\Sigma^\a_L$ and the true edge operators ${\tilde \Sigma}^\a_L$.

To obtain the time constant $T_2^*$, we first disorder average the quantity $\langle \Sigma^z_L(t) \rangle$ to obtain a smoothly oscillating curve. We found that the envelope of its decay fit well to a stretched exponential, of the same form as before, at early times. We extract from this a $T^{*}_{2}$ decay time which does not scale exponentially with system size.

\begin{figure}[tbh]
\includegraphics[width=1.0\linewidth]{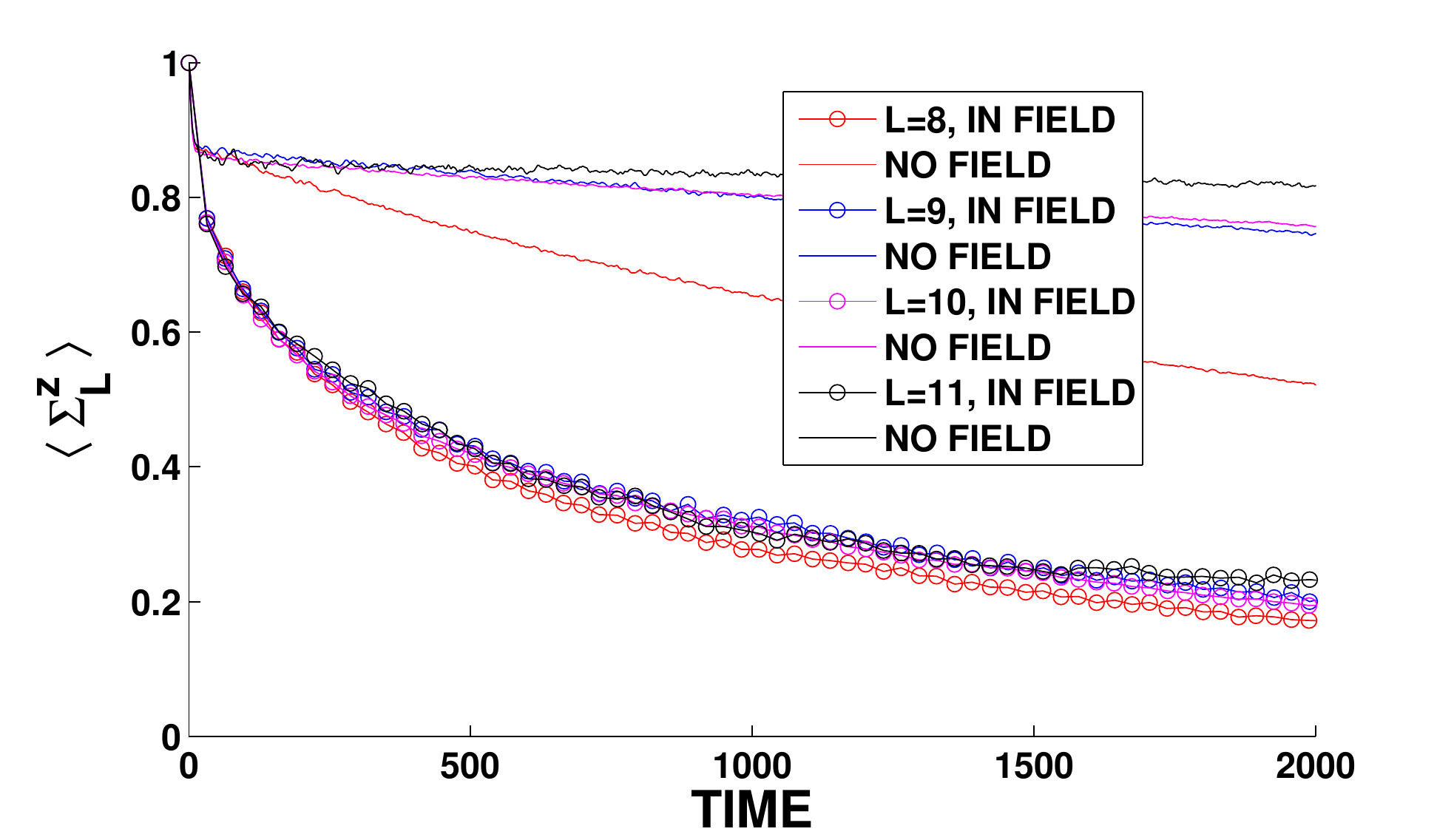}
\caption{Decay of the oscillations of the edge spin subject to an edge field on the time scale $T_2^*$ for varying system length compared to the decay of the spin at zero field $T_0$.  Parameters used had zero mean and standard deviations were $(\sigma_\lambda,\sigma_V,\,\sigma_h)=(1.0,\,0.1,\, 0.05)$.
 }
\label{fig:t2s}
\end{figure}

\section{Edge spin echo -- $T_2$ time}

To obtain the decay of the spin echo as a function of the field reversal time we studied sample sizes L=7-11, with averages over $\{ 8000, 5000, 5000, 2000, 3000 \}$ samples, respectively, for intermediate values of the echo reversal time. We considered the Hamiltonian $H_{0} \pm B \Sigma^{x}_{L}$ before and after the echo, with $B=0.05$, starting from a state initially polarized along +z. The data fit well to a stretched exponential of the form $\left[ \langle \S^z_L(2*T_{R}) \rangle \right]_{dis} \sim C_{1} e^{-\sqrt{\frac{T_{R}}{T_{2}}}}$.

\begin{figure}[tbh]
\includegraphics[width=1.0\linewidth]{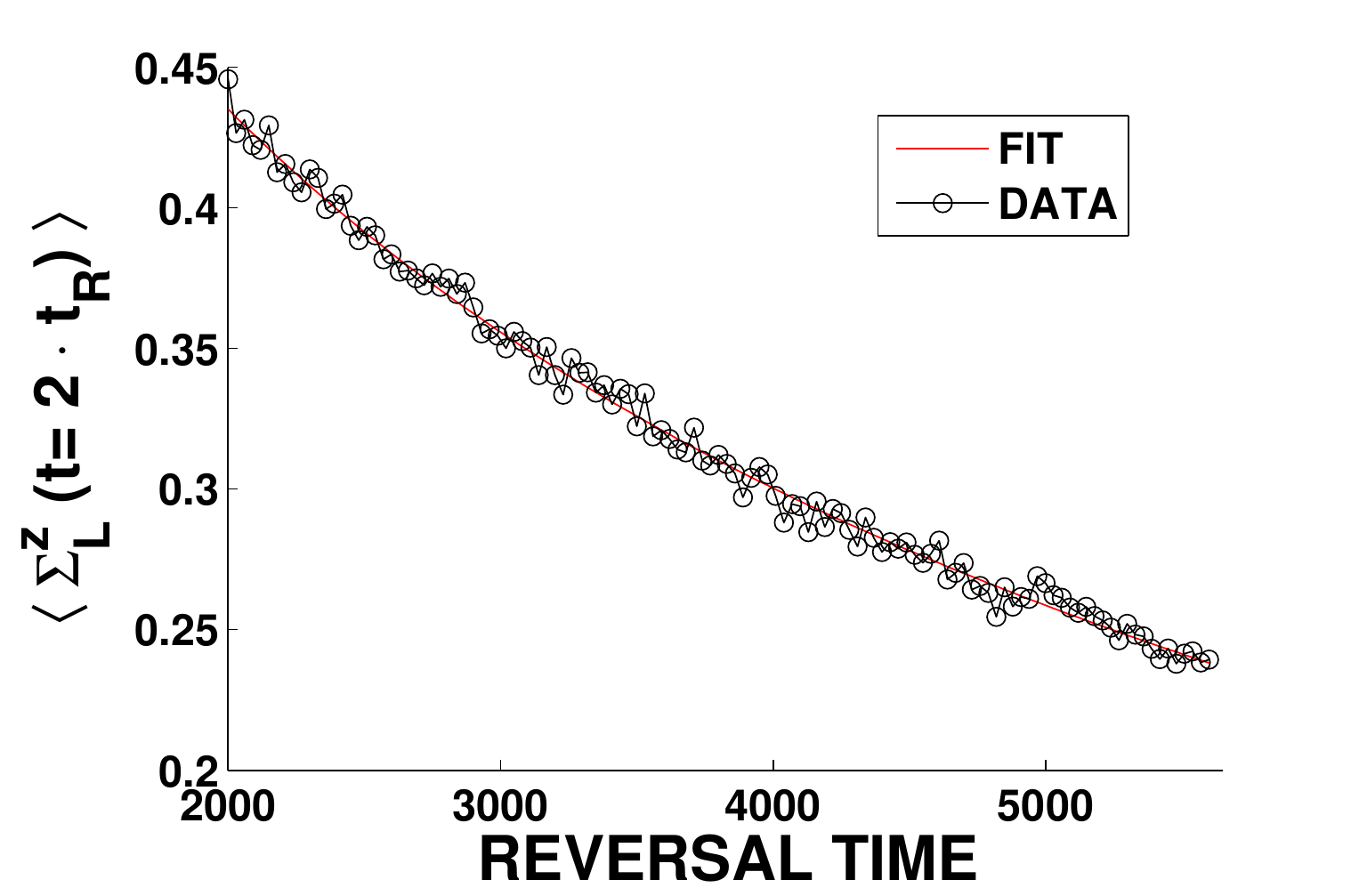}
\caption{The disorder averaged echo signal degrades over a much longer timescale $T_2$, which depends on the system length (L=7 shown).  Parameters used had zero mean and std. deviations were $(\sigma_\lambda,\sigma_V,\,\sigma_h)=(1.0,\,0.1,\, 0.05)$. }
\label{fig:t2}
\end{figure}

\section{Entanglement spectrum}

In the main text, we discussed the survival of edge coherence as a result of the topological nature of the entire spectrum. Here we demonstrate the topological character of high energy states in a system with periodic boundary conditions by considering the entanglement spectra. The entanglement spectra are obtained by partitioning the system into two equal parts and computing the Schmidt values $\lambda_i$. The entanglement spectrum is given by $\epsilon_i=-\ln(\lambda_i)$.

\begin{figure}[t]
 \subfigure{(a)} {\includegraphics[width=1.0\linewidth]{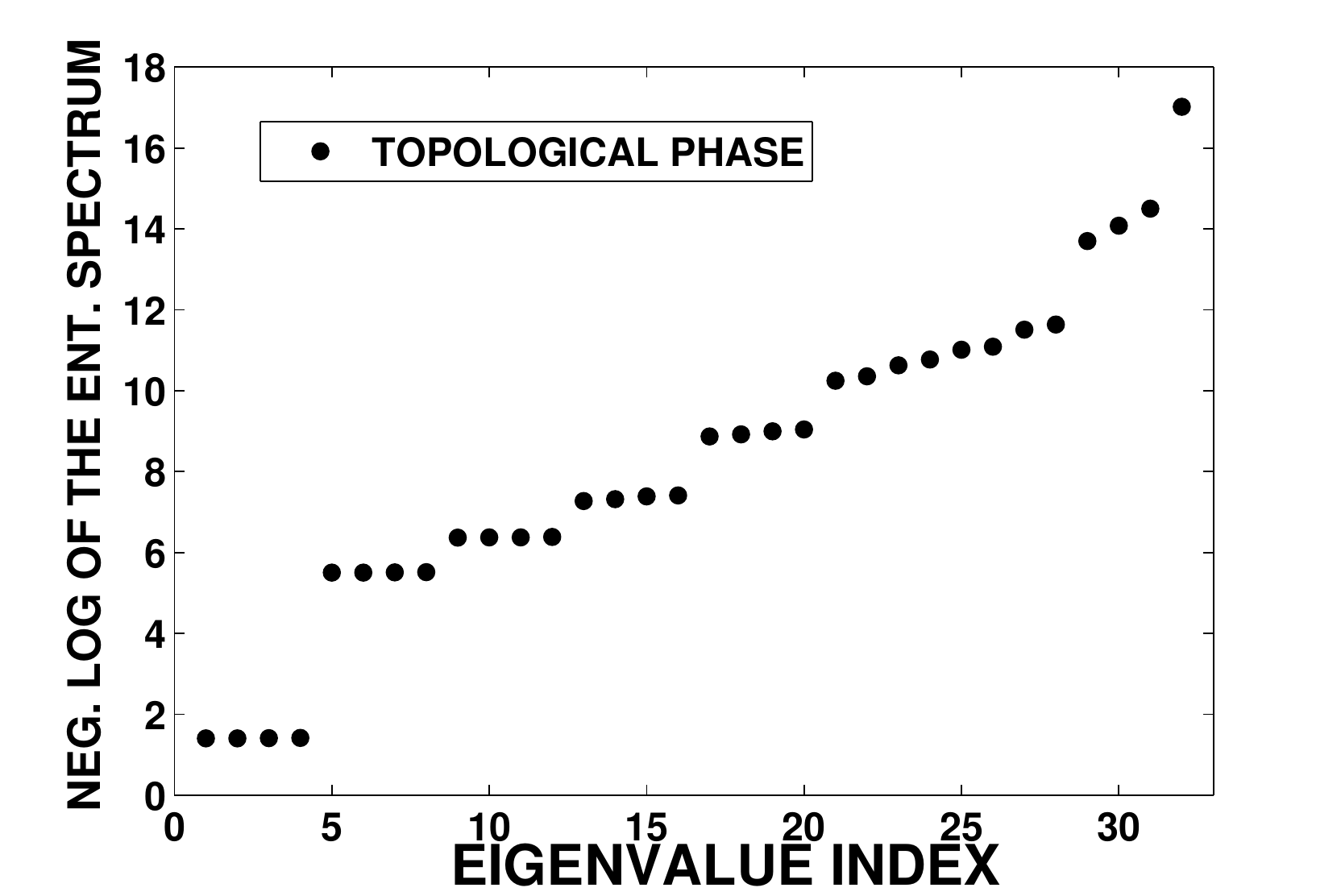}}
  \subfigure{(b)} {\includegraphics[width=1.0\linewidth]{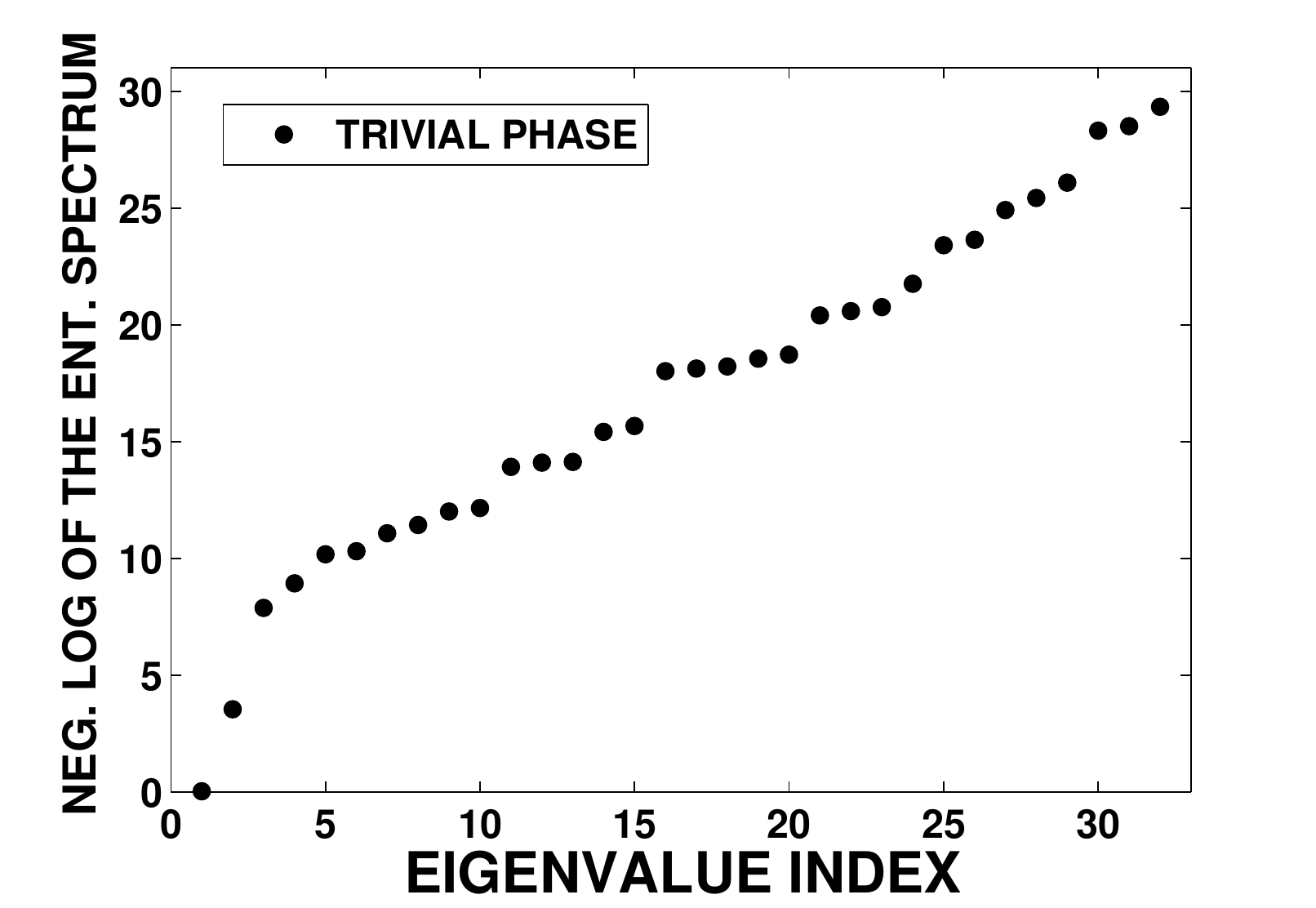}}
\caption{Entanglement spectra in eigenstate number 500 of particular disorder realizations corresponding to the topological state (top) and the trivial state (bottom).}
\label{fig:ent}
\end{figure}

We compute the entanglement spectra for a system of size $L=10$ sites long taking an eigenstate number $500$ approximately in the middle of the spectrum.
Fig. \ref{fig:ent} contrasts the entanglement spectrum of the excited state in the topological state and the trivial state. The four-fold degeneracy in the topological phase corresponds to the spin half degrees of freedom released at each end of the cut. Note, finite size effects are expected to become more important for higher entanglement energy Schmidt states.

\section{Spin Echo}

We consider the time evolution of our system in the presence of an edge field acting on the ``bare" edge spin $H(g)=H_0+g\S^x_L=H_0+g\s^x_1 \s^z_2$.
In the initial state
\be
\ket{\psi_0}=\ket{\ua\ua\ldots\ua}
\ee
the bare edge spin points along $\S^z_L=\s^z_1$. We eventually want to work in the eigenbasis of the generic interacting Hamiltonian in weak field. We choose to proceed in steps by first relating to the interacting zero-field problem. While we are in the localized topological phase, we can label eigenstates with quantum numbers corresponding to the value of the edge spin and bulk conserved operators. We write the zero-field interacting basis as $\ket{\ts,\ta,0}$, which has edge spin ${\tilde\S}^x_L$ equal to $\tilde \s =\pm 1$ and bulk conserved operators ${\tilde K}_i$ equal to $\{\tilde{\kappa}^\a_i\}$; it is continuously connected to the ``bare" $V,h=0$ limit $\ket{\s,\a,0}$, where $0$ denotes zero field. Our initial state is a near perfect eigenstate of $\tilde{\S}^{z}_L$ when $V, h$ are small. The two bases are related by the overlap:

\be
\begin{split}
\bra{\tilde \s', \tilde \a', 0} \s,\a,0 \rangle &=B_{\s\a}(V,h)\d_{\s \s'}\d_{\a \a'}+\\
& \mathcal{O}(V,h)(1-\d_{\s \s'}\d_{\a \a'})
\label{overlap0}
\end{split}
\ee

\noindent where $B_{\s\a}$ has the form $(1-\mathcal{O}(V^{2},h^{2},Vh))e^{i\beta(V,h)}$. Here we took $V$ and $h$ to be uniform on the chain for simplicity. In general they can be disordered as well. Then, the above result should be understood as giving the change of the overlaps with the typical values of $V$ and $h$ on the chain. As we shall see below, the diagonal elements $B_{
\s\a}$ will contribute to the echo, whereas the off-diagonal elements will give an incoherent contribution. From here on, we only keep components of the state which will contribute to the coherent part:
\be
\begin{split}
\ket{\psi_0}&=\sum_{\a} \eta_\a \sum_{\s} \ket{\s,\a,0} \\
	        & \rightarrow \sum_{\a} \eta_\a \sum_{\s} B_{\s\a}(V,h) \ket{\tilde{\s},\tilde \a,0}
\end{split}
\ee

Note that $\eta_{\a}$ is independent of edge spin. The time evolution, on the other hand, is simple in the basis of $H(g)$ eigenstates. The many-body localized topological phase persists in weak fields, so we can label eigenstates of $H(g)$ as $\ket{\tilde\s,\tilde\a,g}$, using the old quantum numbers. For brevity of notation we will sometimes combine the two labels $\tilde{\s}$ and $\tilde{\a}$ to a single label $a$.

The initial state in the basis of $H(g)$ eigenstates is:

\be
\ket{\psi_0}=\sum_{a'} \sum_{a} \eta_{\a'} B_{a'}  \ket{a,g}\bra{a,g}a',0\rangle
\ee

Because the edge field is small there is a strong overlap between the eigenstates with and without a field:
\be
\bra{a,g}a',0\rangle=C_{a}(g,V,h)\d_{aa'}+g\gamma_{aa'}(1-\d_{aa'})
\label{overlap1}
\ee
where all of the off-diagonal elements $\gamma_{aa'}$ are essentially zero
because of localization. Only a restricted set of states can be connected to a given $\ket{a,0}$ by the local edge field, namely those whose local quantum numbers differ within the dimensionless distance $\xi$ of the edge.  The number of such states scales exponentially with the localization length. Imposing normalization in (\ref{overlap1}) to lowest order in $g$ gives $C_{a}(g,V,h) = (1-c_{a}(\xi,V,h) g^{2}) e^{i\psi_{a}}$, where $c_{a}(\xi,V,h)$ is exponential in $\xi$ and vanishes as $V, h \rightarrow 0$. The phase $\psi_a$ can be (and in general is) first order in $g, V, h$. As we shall see, only the first term in (\ref{overlap1}) will lead to a coherent echo in the evolution following field reversal.

After time evolution, the contribution is
\be
\ket{\psi(t)}=\sum_{\s\a}e^{-i E_{\s\a} (g)t} \eta_{\a} B_{\s\a}C_{\s\a}\ket{{\tilde \s},\tilde{\a},g}
\label{psit}
\ee

We want to run this time evolution up to the reversal time $t_R$, at which point we invert $g$.
The ensuing dynamics is simple when written in terms of the eigenstates of $H(-g)$. Therefore to track this part of the time evolution we convert to the basis $\ket{{\tilde\s}, \tilde{\a},-g}$
\be
\begin{split}
\ket{\psi(t)}=\sum_{a}\sum_{a'} &e^{-iE_a(g) t} \eta_{\a} B_a C_a\\
&\ket{a',-g}\langle{a',-g}\ket{a,g}
\end{split}
\ee
This is similar to the conversion which we described above from the $H_0$ basis to the $H(g)$ basis. Again there is a strong overlap between the states
\be
\langle{a',-g}\ket{a,g}= D_{a}(g,V,h) \d_{aa'}+ g \chi_{aa'}(g) (1-\d_{aa'}).
\label{overlap2}
\ee
and by the same token we have, to lowest order in $g$, $D_{a}(g,V,h)= (1-d_{a}(\xi,V,h) g^{2})e^{i\zeta_{a}}$. Only the diagonal term in (\ref{overlap2}) will contribute to a coherent echo.


Next, we note that the spectrum of $H_0$ on a semi-infinite system is doubly degenerate $E_{\s\a}(0)=E_\a$. We will follow the evolution of a doublet with fixed $\{\tilde{\k}^{\a}_i\}$ and $\tilde \s=\pm 1$ upon increasing $g$ to positive or negative values:
\bea
\D E_{\s,\a}(g)&=&f_1(\tilde \s,\tilde{\a}) g+ f_2(\tilde \s,\tilde{\a}) g^2\nn\\
&&+ f_3(\tilde \s,\tilde{\a}) g^3+\ldots
\label{DE}
\eea
The form of this energy splitting is constrained by symmetry.
The unitary operator $U=\prod_i\s^x_{2i}$, one of the $Z_2$ symmetry generators, transforms $H(g)\to H(-g)$, which means that if $H(g)\ket{\psi}=E\ket{\psi}$ then $H(-g)(U\ket{\psi})=E(U\ket{\psi})$. Furthermore, since we label the eigenstates of $H(g)$ with the quantum numbers of the integrals of motion of $H_0$, we can write the transformation of the eigenstates as
\be
U\ket{\tilde \sigma, \tilde{\a},g}=e^{i\psi_\a} \ket{-\tilde \sigma, \tilde{\a},-g}.
\label{Upsi}
\ee
Putting together the transformation of the energies and of the eigenstate labels under $U$ we have
\be
\D E_{\s,{\a}}(g)=\D E_{-\s,\a}(-g).
\ee
Therefore all even indexed functions $f_{2i}$ appearing in the expansion (\ref{DE}) must be independent of $\tilde \s$, whereas all odd indexed functions must be proportional to $\tilde \s$. We may therefore write this expansion more simply as
\bea
\D E_{\s,\tilde{\a}}(g)&=&f_1(\tilde{\a})\tilde \sigma g+ f_2(\tilde{\a}) g^2\nn\\
&&+ f_3(\tilde\a) \tilde \sigma g^3+\ldots
\eea
We now run the time evolution using the above expressions for $\D E_{\s,\a}(g)$ up to the  time $t_R$ and from there up to time $t=2t_R$ with $g\to -g$. The result is
\bea
\ket{\psi_{\text{echo}}}&=&\sum_\a \eta_{\a} e^{i\phi_\a}
 \Big(B_{\ua\a}C_{\ua\a}D_{\ua\a} \ket{\tilde \ua,\tilde{\a},-g}\nn\\
 &&~+B_{\da\a}(g) C_{\da\a}D_{\da\a} \ket{\tilde\da,\tilde{\a},-g}\Big)
\eea
Crucially the contribution of the odd terms in the expansion of $\D E$ accumulated in the evolution up to the reversal time is exactly cancelled in the subsequent evolution with $g\to -g$.  $\phi_\a$ is a random phase {\em independent} of $\s$ which is accumulated due to the even terms in the expansion including the zeroth order energy $E_\a(g=0)$.

In order to measure the spin in the final state we make a third and last conversion, using (\ref{overlap1}), from the $\ket{\SA,-g}$  back to the basis $\ket{\SA,0}$. The final state coherent contribution to the echo will come from:
\be
\begin{split}
&\ket{\psi_{\text{echo}}}=\sum_\a \eta_{\a} e^{i\phi_\a}\Big(R_{\ua\a}\ket{\UA,0}
+R_{\da\a}\ket{\DA,0}\Big)\nn\\
&R_{\s\a}\equiv\\
&B_{\s\a}(V,h)C_{\s\a}(-g,V,h)^*D_{\s\a}(g,V,h)C_{\s\a} (g,V,h)
\end{split}
\ee
Using the the transformation of the eigenstates (\ref{Upsi}) we can relate $R_{\ua\a}$ and $R_{\da\a}$. Inserting the $U\yd U$ into each overlap that makes up $R_{\da\a}$ and momentarily dropping the fixed $\a$ label, we get:
\be
\begin{split}
&R_{\da\a}\\
&=\bra{\tda,0}\tda,-g\rangle\bra{\tda,-g}\tda,g\rangle\bra{\tda,g}\tda,0\rangle\bra{\tda,0}\da,0\rangle\\
&=\bra{\tua,0}\tua,g\rangle\bra{\tua,g}\tua,-g\rangle\bra{\tua,-g}\tua,0\rangle\bra{\tua,0}\ua,0\rangle\\
&=\left(\bra{\ua,0}\tua,0\rangle\bra{\tua,0}\tua,-g\rangle\bra{\tua,-g}\tua,g\rangle\bra{\tua,g}\tua,0\rangle\right)^*\\
&=R_{\ua\a}^*\equiv R^*_\a
\end{split}
\ee
Hence we can write:
\be
\ket{\psi_{\text{echo}}}=\sum_\a \eta_\a e^{i\phi_\a}\Big(R_{\a}\ket{\UA,0}
+R_{\a}^*\ket{\DA,0}\Big)
\ee
The expectation value of $\tilde{\Sigma}^z_L$ in the final state is then given by:
\be
\av{\tilde\Sigma^z_L}=\sum_\a |\eta_\a|^{2} \Re\left[R_\a(g,V,h)^2\right]
\ee
Like the elements that compose it, $R_{\s\a}$ has, to lowest order, the general form:
\be
\begin{split}
&R_{\da\a}(g,V,h)=R_{\ua\a}^*(g,V,h)\\
&=\left(1 - r_{1\a}(\xi,V,h) g^{2} - r_{2\a}V^{2} - r_{3\a} Vh - r_{4\a} h^{2}\right) e^{i\t_\a(g,V,h)}
\end{split}
\ee

where $\t_\a$ in general may start at linear order in its variables and $r_{1\a}$ is exponential in $\xi$. Substituting this in the expectation value of the edge spin we find
\be
\av{\tilde{\S}_L^z}\approx 1-\mathcal{O}(Vg^{2},h g^{2},V^{2},Vh,h^{2})
\ee

The $g$ corrections are linked with $V, h$ corrections, since we expect a perfect echo in the $V,h \rightarrow 0, g \neq 0$ limit. There is a final factor left out, from the fact that we measure a ``bare" operator, the strictly local operator $\S^z_L$ rather than the true quasi-local edge spin $\tilde{\S}^z_L$ of the system with interaction $V$. This extra suppression is of the same origin as $B(V,h)$ and is similar in form, $(1-\mathcal{O}(V^{2},Vh,h^{2}))$.

Thus, deep in the topological phase we recover the expectation value $\av{\Sigma^z_L}$ in the final state up to the corrections of order $\mathcal{O}(Vg^{2},h g^{2},V^{2},Vh,h^{2})$. Even when these terms lead to a strong suppression, the echo will be non-zero in the limit $t_R\to\infty$ as long as the bulk is in the topological phase.



\end{document}